\documentclass[journal=apchd5,manuscript=article]{achemso}

\usepackage[version=3]{mhchem} 
\usepackage{xcolor}

\author{Diego Martin-Cano}
\email{diego-martin.cano@mpl.mpg.de}
\author{Harald R. Haakh}
\altaffiliation{Past address}
\affiliation[MPL]
{Max Planck Institute for the Science of Light, Staudtstr. 2, D-91058 Erlangen, Germany}
\author{Nir Rotenberg}
\email{nir.rotenberg@nbi.ku.dk}
\affiliation[NBI]
{Niels Bohr Institute and Center for Hybrid Quantum Networks (Hy-Q), University of Copenhagen, Blegdamsvej 17, DK-2100 Copenhagen, Denmark}

\title[Chiral Emission]
  {Chiral emission into nanophotonic resonators}

\abbreviations{IR,NMR,UV}

\begin{document}
\begin{abstract}
 Chiral emission, where the handedness of a transition dipole determines the direction in which a photon is emitted, has recently been observed from atoms and quantum dots coupled to nanophotonic waveguides.  Here, we consider the case of chiral light-matter interactions in resonant nanophotonic structures, deriving closed-form expressions for the fundamental quantum electrodynamic quantities that describe these interactions.  We show how parameters such as the position dependent, directional Purcell factors and mode volume can be calculated using computationally efficient two dimensional eigenmode simulations.  As an example, we calculate these quantities for a prototypical ring resonator with a geometric footprint of only 4.5~$\mu$m$^2$, showing that perfect directionality with a simultaneous Purcell enhancement upwards of 400 are possible.  The ability to determine these fundamental properties of nanophotonic chiral interfaces is crucial if they are to form elements of quantum circuits and networks.
\end{abstract}

\begin{figure*}[t!]
\begin{center}
\includegraphics[width=8cm]{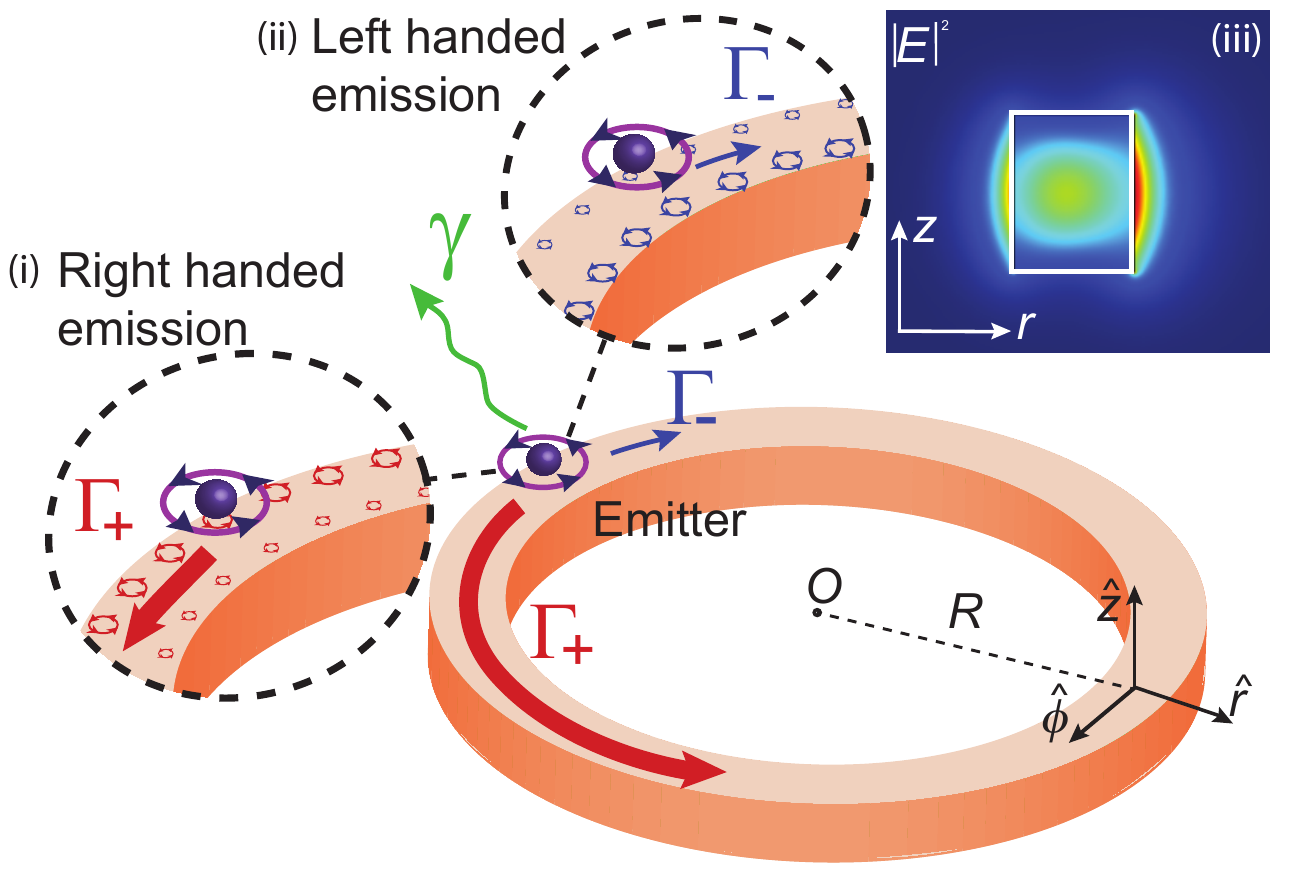}
\caption{ Chiral emission into a whispering-gallery-mode photonic resonator.  The main figure sketches the emission of a right-handed transition dipole located on the outer edge of a ring resonator, where a fast emission rate into the right-handed mode $\left(\Gamma_+\right)$ and slow emission rate into the left-handed mode $\left(\Gamma_-\right)$ are shown with a thick red line and a thin blue line, respectively. All other losses are characterized by the rate $\gamma$.  The asymmetric emission occurs because, at the position of the emitter, the polarization of the electric field of the right-handed mode is primarily right-handed (and hence overlaps well with the dipole vector, inset i.). Conversely, the field of the left-handed mode, at the position of the emitter, is primarily left-handed polarized (inset ii.).  The size of the circles, in insets i. and ii., indicate the fraction of the electric field that is right-handed polarized and would therefore overlap with the transition dipole moment. For completeness, inset iii. shows the amplitude of the electric field distribution of a ring resonator with radius $R = 1.19 \, \mu$m, 162 nm width and 210 nm height (refractive index $n=3.2$ of  gallium phosphide, surrounded by air), with maxima observed both inside and outside the waveguide.}
\label{fig:f1_system}
\end{center}
\end{figure*}

The emergence of chiral light-matter interactions in nanophotonic systems has redefined our understanding of fundamental processes such as the emission or scattering of photons by quantum emitters.\cite{Lod17}  In nanophotonic structures the spin and momentum of a local light fields can be locked~\cite{Bli14}, meaning that at nanoscopic length scales the direction in which the orientation vector of the light field rotates depends on the direction in which photons travel.  As a direct consequence, counter-propagating modes will interact differently with emitters whose transitions are described by elliptically polarized transition dipoles, resulting in highly unidirectional emission or scattering as sketched in Fig.~\ref{fig:f1_system}.  Chiral interactions have recently been observed in a variety of classical\cite{Pet14, Feb15} and quantum\cite{Mit14, Soe15, Col16} nanophotonic systems, including in optical nanofibers, nanobeams and photonic crystal waveguides.  The existence of these chiral quantum elements has motivated a number of recent proposals that include new routes toward robust generation of entanglement\cite{Ram14,Gon15}, nonreciprocal circuits\cite{Soe15, Gon16} and quantum processing with chiral networks.\cite{Pic15, Mah16}  Moreover, our growing understanding of chiral effects has led to novel technological solutions to outstanding photonic and quantum optical challenges, such as an optical isolator at the single-photon level\cite{Sch16} and the demonstration of robust, path-dependent preparation of the spin of a single quantum dot\cite{Col17}.

A cavity-based chiral quantum interface would direct photons from a quantum emitter primarily into one of two counter-propagating modes, while enhancing this chiral light-matter interaction.  In a whispering gallery mode resonator (WGM), such as the ring illustrated in Fig.~\ref{fig:f1_system}, this means that the rate of emission into the left-handed ($\Gamma_-$) and right-handed ($\Gamma_+$) degenerate modes differ.
This, in turn, means that the total probability of emitting a single photon into each mode $\left(\beta_\pm\right)$ is modified, since
\begin{equation}\label{eq:beta}
  \beta_\pm = \frac{\Gamma_\pm}{\Gamma_+ + \Gamma_- + \gamma},
\end{equation}
where $\gamma$ is the rate at which light couples to all other modes via radiative and nonradiative processes.  The viability of a chiral quantum interface is to a large extent determined by the losses, which ideally should be very small: $\gamma \ll \Gamma_+ + \Gamma_-$.  In practice, this requirement can be met either by suppressing radiative losses through careful engineering of waveguides\cite{Arc14} or, increasingly, through the enhancement of emission into the desired modes using resonant structures\cite{Sch17, Rot17}.  The latter approach shows much promise, particularly with solid-state emitters, as it can compensate for radiation losses and dephasing due to the phononic environment, and provide higher photon count rates.\cite{Wan16}  Historically, this type of enhancement of the emission of linear dipoles has been characterized by the Purcell factor, $F = \Gamma / \Gamma_0$, where $\Gamma$ is the decay rate into a single resonant mode and $\Gamma_0$ is the emission rate into a homogenous material hosting the emitter.\cite{Pur46}  Finding an analogous quantum electrodynamical quantity for chiral emission is critical if we are to understand the ultimate limits of chiral quantum effects and devices.  Here we address this need, providing closed-form expressions for $\Gamma_\pm$ and $F_\pm$ for nanophotonic resonators and, in the process, introducing the concept of a chiral mode volume $V_\pm$.

\section{Formalism}
\subsection{Chiral nanophotonic resonators}
As is the case for linear waveguides, a resonator-based chiral quantum interface demands the presence of two independent, counter-propagating modes whose complex frequencies are degenerate $\omega_+ = \omega_- = \tilde{\omega}$.\bibnote{Here, the real part of $\tilde{\omega}$ is the resonance frequency while the imaginary component quantifies the damping of the resonator.}  In practise, this precludes the use of Fabry-Perot cavities, rather requiring the use of WGM resonators such as bottle, disk or ring microstructures. These support two degenerate quasi-normal modes (QNM) with opposite propagation directions, which are solutions of the inhomogeneous Maxwell Equations. These QNMs, so-called since they can experience both radiation and absorption losses\cite{Lal17}, can be expressed  by electric and magnetic field distributions, e.g.
\begin{equation}\label{eq:ModeSep}
  \tilde{\mathbf{E}}_\pm \left(\mathbf{r}\right) = \bar{\mathbf{E}}_\pm\left(r, z\right)e^{\mp iq\varphi}.
\end{equation}
Here, we have explicitly accounted for the  symmetry of the microresonators under consideration by separating the longitudinal (i.e. the azimuthal variable $\varphi$, see Fig.~\ref{fig:f1_system}) dependency from the transverse one, noting that the mode must fit an integer number $q$ of waves into the resonator.  A prototypical transverse intensity distribution of a QNM $\left|\bar{\mathbf{E}}_+\left(r, z\right)\right|^2$ is illustrated in the inset of Fig.~\ref{fig:f1_system}, where the confinement of the light to the high-index guiding region is clearly visible.  Time reversal symmetry of Maxwell's Equations ensures that $\tilde{\mathbf{E}}_+ = \tilde{\mathbf{E}}_-^*$\cite{Lod17}, where the asterisk indicates complex conjugation. For the WGM resonators this results in the following relations for the components $\bar{\mathbf{E}}_\pm\left(r, z\right)$:\cite{Col90}
\begin{alignat}{3}\label{eq:Ep2m}
  \nonumber & \bar{E}_{r, +} = -\bar{E}_{r, -} \quad && \bar{E}_{z, +} = -\bar{E}_{z, -}\, \quad && \bar{E}_{\varphi, +} = \bar{E}_{\varphi, -}\\
  & \bar{H}_{r, +} = \bar{H}_{r, -} \quad && \bar{H}_{z, +} = \bar{H}_{z, -}\, \quad && \bar{H}_{\varphi, +} = -\bar{H}_{\varphi, -}
\end{alignat}
The change of sign of the transverse electric field components ensures that the polarization ellipse traced out by the $E_r$ and $E_\varphi$ components of the counter-propagating QNMs rotate in the opposite directions\bibnote{Note that the same is true for the ellipse of the $E_z$ and $E_\varphi$ components, and for the magnetic components, where sign of the longitudinal component flips}.  With these properties of the QNMs of the WGM resonator, we can now set out to derive expressions for $\Gamma_\pm$, $V_\pm$ and, ultimately, $F_\pm$.

\subsection{General expressions for chiral emission rates}
We use the first-principle approach introduced by Sauvan \emph{et al.},\cite{Sau13} which is based on the QNM expansion of Maxwell's Equations and Fermi's golden rule. Here the electromagnetic field emitted by an electric dipole in a nanophotonic system is expanded as a sum of QNMs,
\begin{equation}\label{eq:QNM}
  \mathbf{E}\left(\mathbf{r},\omega\right)\approx\sum_{m=1}^{M}\alpha_{m}\left(\omega\right)\tilde{\mathbf{E}}_{m}\left(\mathbf{r}\right),
\end{equation}
where $\alpha_{m}\left(\omega\right)$ are the complex coefficients that describe the amplitude and relative phase that each QNM contributes. A similar expression may be written for the magnetic field.

Each QNM fulfills the Lorentz Reciprocity Theorem,
\begin{eqnarray}
  \nonumber -\omega\mathbf{p}\cdot\tilde{\mathbf{E}}_{n}\left(\mathbf{r}_{0}\right)&=& \int d^{3}\mathbf{r}\left\{ \mathbf{E}\cdot\left[\omega\varepsilon\left(\omega\right)-\tilde{\omega}_{n}\varepsilon\left(\tilde{\omega}_{n}\right)\right]\tilde{\mathbf{E}}_{n} \right. \\
   \label{eq:Reciprocity}
   &-&
   \left. \mathbf{H}\cdot\left[\omega\mu\left(\omega\right)-\tilde{\omega}_{n}\mu\left(\tilde{\omega}_{n}\right)\right]\tilde{\mathbf{H}}_{n}\right\},
\end{eqnarray}
that relates the electric field radiated by a dipole moment $\mathbf{p}$ located at position $\mathbf{r}_{0}$ into one QNM at frequency $\omega_n$ to the overlap between that QNM and the total electromagnetic field at frequency $\omega$.  Here, $\varepsilon=\epsilon_{0}\varepsilon_{r}$ and $\mu=\mu_{0}\mu_{r}$ are the permittivity and permeability characterizing the inhomogeneous nanophotonic system. These magnitudes vary slowly as a function of frequency over the typical narrow bandwidths of microresonators and emitters, so that dispersion is neglected.\bibnote{Note that a generalization to dispersive resonators is nevertheless possible\cite{Sau13}}
Eq.~\ref{eq:QNM} can be used to reformulate Eq.~\ref{eq:Reciprocity} as a linear system of equations,
\begin{equation}\label{eq:LinSys}
  \sum_{m}A_{nm}\left(\omega\right)\left(\omega-\tilde{\omega}_{m}\right)\alpha_{m}\left(\omega\right)=-\omega\mathbf{p}\cdot\tilde{\mathbf{E}}_{n}\left(\mathbf{r}_{0}\right),
\end{equation}
where
\begin{equation}\label{eq:Amn}
  A_{nm}\left(\omega\right)=\frac{\left(\omega-\tilde{\omega}_{n}\right)}{\left(\omega-\tilde{\omega}_{m}\right)}\int d^{3}\mathbf{r}\left[\tilde{\mathbf{E}}_{m}\cdot\varepsilon\tilde{\mathbf{E}}_{n}-\tilde{\mathbf{H}}_{m}\cdot\mu\tilde{\mathbf{H}}_{n}\right].
\end{equation}
Solving these equations yields the coefficients $\alpha_m\left(\omega\right)$ of the QNM expansion of Eq.~\ref{eq:QNM}.

In obtaining these, we diverge from the work of Sauvan et al., who treated the case of normal emission from linear dipoles~\cite{Sau13} into individual  non-degenerate modes.  In contrast, the asymmetric chiral emission considered here requires the explicit treatment of complex-valued dipole vectors and  degenerate counter-propagating modes, meaning that both coefficients, $\alpha_+$ and $\alpha_-$, must be calculated. \bibnote{Assuming that the modes do not spectrally overlap, we need only consider the two counterpropagating modes so that the sum runs over $\{+,-\}$ and we omit the frequency index for brevity; the calculation can be repeated for each pair of degenerate QNMs.} In this $\left(+, -\right)$ basis, $A_{++} = A_{--} = 0$ due to the azimuthal symmetry of the QNM (Eq.~\ref{eq:ModeSep}) and the volume integral in Eq.~\ref{eq:Amn}.  Consequently, we can write
\begin{equation}\label{eq:AmnSolution}
  A_{+-}(\omega) =
   A_{-+}\left(\omega\right)
  =\int d^{3}\mathbf{r}\left[\tilde{\mathbf{E}}_{+}\cdot\varepsilon\tilde{\mathbf{E}}_{-}-\tilde{\mathbf{H}}_{+}\cdot\mu\tilde{\mathbf{H}}_{-}\right].
\end{equation}
This expression is used to invert $\mathbf{A}$ and applied in conjunction with Eq.~\ref{eq:LinSys} to obtain
\begin{eqnarray}
\alpha_{\pm} = - \frac{\omega \mathbf{p}\cdot \mathbf{E}_{\mp}(\mathbf{r}_0)}{(\omega- \tilde \omega)A_{+-}(\omega)}.
\end{eqnarray}
Fermi's Golden Rule can relate the total decay rate to contributions of the individual QNMs through
\begin{equation}\label{eq:FGR}
  \Gamma =\frac{2}{\hbar}\mbox{Im}\left[\sum_m\alpha_{m}\mathbf{p}^{*}\cdot\tilde{\mathbf{E}}_{m}\left(\mathbf{r}_{0}\right)\right],
\end{equation}
so that the emission rate into each of the counter-propagating modes is
\begin{equation}\label{eq:Gamma}
  \Gamma_{\pm} = -\frac{2}{\hbar}\omega\mbox{Im}\left\{ \frac{\left[\mathbf{p}\cdot\tilde{\mathbf{E}}_{\mp}\left(\mathbf{r}_{0}\right)\right]\left[\mathbf{p}^{*}\cdot\tilde{\mathbf{E}}_{\pm}\left(\mathbf{r}_{0}\right)\right]}{\left(\omega-\tilde{\omega}\right)\int d^{3}\mathbf{r}\left[\tilde{\mathbf{E}}_{+}\cdot\varepsilon\tilde{\mathbf{E}}_{-}-\tilde{\mathbf{H}}_{+}\cdot\mu\tilde{\mathbf{H}}_{-}\right]}\right\}.
\end{equation}
This compact equation captures the physics of chiral emission since it enables asymmetric directional coupling, i.e. $\Gamma_+ \neq \Gamma_-$:  if, for example, there is a large overlap between $\mathbf{p}^{*}$ and $\tilde{\mathbf{E}}_{+}$ such that $\Gamma_{+}$ is maximized, there will be little overlap between $\mathbf{p}^{*}$ and $\tilde{\mathbf{E}}_{-}$ and thus $\Gamma_{-}$ will be slow.  It is interesting to note that while $\Gamma_{+}$ (and $\Gamma_{-}$) depends on both $\tilde{\mathbf{E}}_{+}$ and $\tilde{\mathbf{E}}_{-}$, the factor with which each mode modifies the emission rate differs for elliptical dipoles.

Equation~\ref{eq:Gamma} can be normalized to the emission of an equivalent dipole into bulk media of refractive index $n$ $\Gamma_0 = \omega^{3}\left|p\right|^{2}n / \left(3\pi\epsilon_{0}\hbar c^{3}\right)$, giving the on-resonance chiral emission enhancement due to the WGM resonator, i.e. the directional Purcell factors
\begin{equation}\label{eq:chiralF}
  F_{\pm}=\frac{\Gamma_{\pm}}{\Gamma_{0}} =\frac{3}{4\pi^{2}}\left(\frac{\lambda_{0}}{n}\right)^{3}\mbox{Re}\left(\frac{Q}{V_{\pm}}\right).
\end{equation}
Here, $\omega_0 = \mbox{Re}\,\tilde{\omega}$ and $Q = \mbox{Re}\,\tilde{\omega} / \left(2\mbox{Im}\,\tilde{\omega}\right)$ are the central frequency and quality factor of the resonance, and we have also defined the position-dependent chiral mode volumes to be
\begin{equation}\label{eq:chiralV}
  V_{\pm}(\mathbf{r}_0)=\frac{\int d^{3}\mathbf{r}\left[\tilde{\mathbf{E}}_{+}\cdot\varepsilon\tilde{\mathbf{E}}_{-}-\tilde{\mathbf{H}}_{+}\cdot\mu\tilde{\mathbf{H}}_{-}\right]}{2\varepsilon\left[\mathbf{u}\cdot\tilde{\mathbf{E}}_{\mp}\left(\mathbf{r}_{0}\right)\right]\left[\mathbf{u}^{*}\cdot\tilde{\mathbf{E}}_{\pm}\left(\mathbf{r}_{0}\right)\right]},
\end{equation}
where $\mathbf{u}$ is the unit vector of the dipole $\left(\mathbf{p} = p \mathbf{u}\right)$. That these mode volumes depend on the position of the emitter might, initially appear odd. Yet, very similarly, the Purcell factor acquires positional dependence  through the local density of optical states if a Green's tensor approach is chosen instead.\cite{Nov12} This position-dependent normalization factor arises directly from our derivation and provides an clear and consistent way to define chiral volumes, particularly for lossy modes and in the presence of absorption, where the usual formulation fails.\cite{Kri15, Lal17}

\section{Resonant chiral emission}

\begin{figure*}[!t]
\begin{center}
\includegraphics[width=0.9\textwidth]{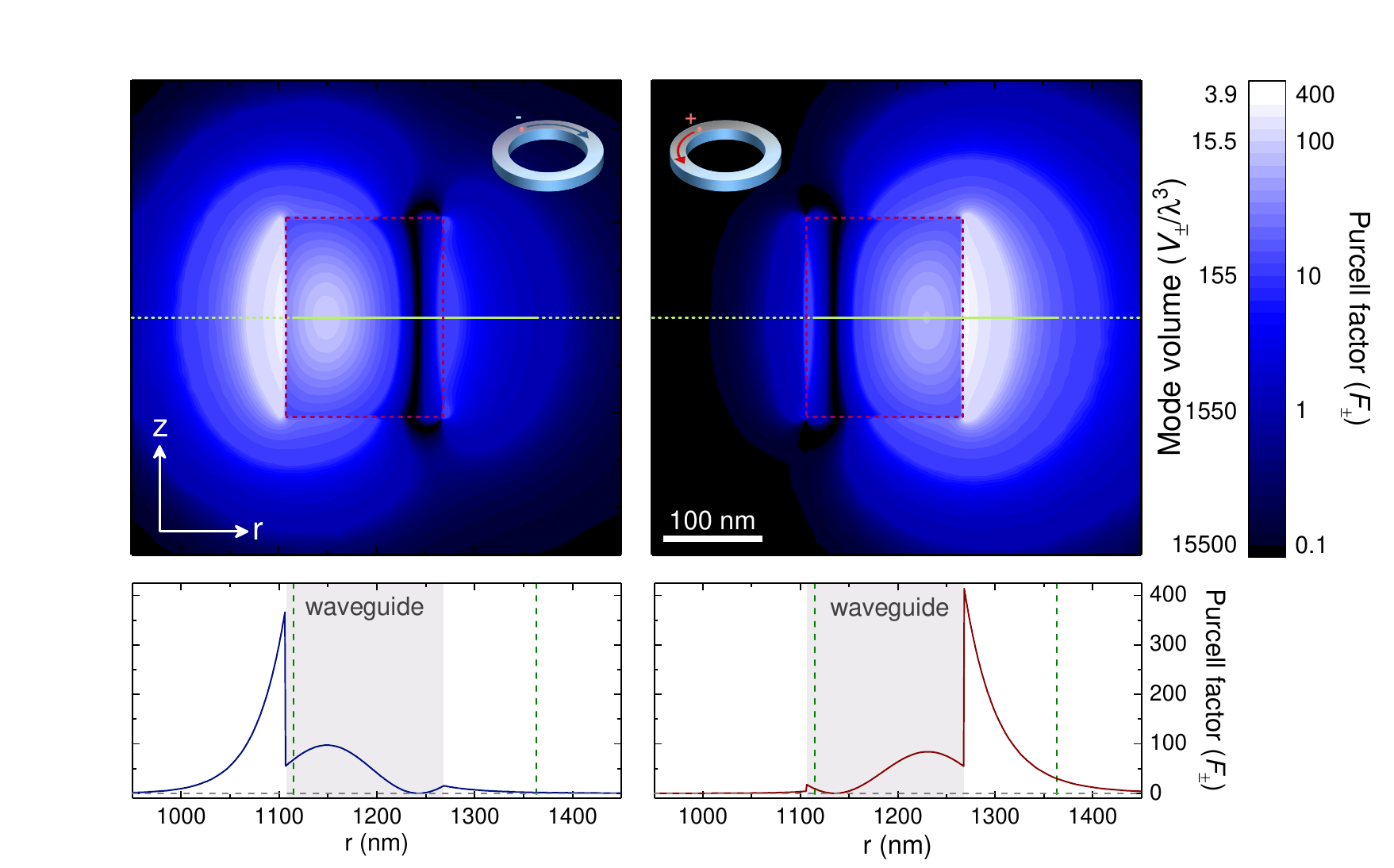}
\vspace{-10px}
\caption{(color online) Spatial dependence of the chiral mode volumes and Purcell factors for emission into the two counter-propagating modes of a ring-resonator,  calculated for a right-hand circular dipole $\mathbf{u}_+ = 1/\sqrt{2}\left(\hat{r} + i\hat{\varphi}\right)$ on a cross-section of the waveguide (dashed square, inner edge to the left).  Emission into the left-handed mode ($\tilde{\mathbf{E}}_{-}$, top left panel) occurs mainly on the inner side of the waveguide, while emission into the right-handed mode ($\tilde{\mathbf{E}}_{+}$, bottom right panel) occurs mainly on the outer side of the waveguide.  For clarity, line cuts of the corresponding Purcell factors along the symmetry axis (dashed line in top panels) are shown in the bottom panels, on a linear scale.  An opposite-handed dipole, $\mathbf{u}_-$, will emit in the opposite fashion (i.e. a swapping of the panels).  The line along which $\beta_{\pm}$ is calculated for Fig.~\ref{fig:f4_beta} below is shown in solid in the top panels, and denoted by the vertical dashed lines in the bottom panels.}
\label{fig:f2_VnF}
\end{center}
\end{figure*}

We now apply our theory to a realistic nanophotonic system that can support and enhance chiral emission, namely the ring resonator sketched out in Fig.~\ref{fig:f1_system}. Due to the azimuthal symmetry of WGM resonators, the QNMs $\tilde{\mathbf{E}}_{\pm}$ can be determined from two-dimensional eigenmode calculations, and so the problem reduces to solving for the different $\alpha_m$'s. As an example, we consider a ring with a radius of only 1.19~$\mu$m, of rectangular cross-section of 210 nm height and 162 nm width, and made out of a gallium phosphide $\left(n = 3.2\right)$ surrounded by air.  For this geometry, the azimuthal mode of order  $g= 21$ is found at a frequency of 393.58 THz, with a moderate $Q \approx 20,000$, well within the range of what can be realistically fabricated.\cite{Bru15}

We calculate the vectorial components of this mode using a finite-element eigenmode solver (Comsol Multiphysics), showing the resultant intensity distribution in the inset of Fig.~\ref{fig:f1_system}.  Equations~\ref{eq:chiralF} and~\ref{eq:chiralV}, in conjunction with the relations between the counter-propagating fields (Eq.~\ref{eq:Ep2m}) allow us to very simply map out the spatial dependence of the chiral mode volumes and Purcell factors, which we show in Fig.~\ref{fig:f2_VnF} for a right-handed circular dipole $\mathbf{u}_+ = \left(\hat{r} + i\hat{\varphi}\right)/\sqrt{2}$.  The difference between $V_+$ and $V_-$, and hence between $F_+$ and $F_-$, is striking: The emitter radiates into the $\tilde{\mathbf{E}}_{+}$ mode when it is placed near the outer (right) side of the bar, while emission into the $\tilde{\mathbf{E}}_{-}$ mode occurs primarily when it is found at the inner (left) side. Note, too, that $V_+$ is not simply a mirror image of $V_-$, as a slight asymmetry is introduced between these two distributions by the bending of the ring, which results in a higher mode intensity and hence emission enhancement at the outer side of the ring.  Changing the dipole polarization rotation to $\mathbf{u}_- = \left(\hat{r} - i\hat{\varphi}\right)/\sqrt{2}$ simply reverses the emission direction (i.e. exchanges the panels of Fig.~\ref{fig:f2_VnF}).

For both modes, we observe that emission peaks first inside the waveguide and then again, with greater amplitude outside, near its interface.  For this structure we calculate a maximal $F_+ \approx 80$ inside and $F_+ \approx 400$ outside the bar, meaning that this specific geometry is highly advantageous for quantum emitters that are embedded in the low-index material surrounding the ring, such as organic molecules in a solid-state matrix or trapped atoms.  Tuning the ring geometry, either by increasing the bar cross-section or the radius of the ring, pushes the mode into the high-index material and hence enhances the emission inside the waveguide, as would be favourable for emitters such as defect centers in diamond or quantum dots.

\begin{figure*}[t!]
\begin{center}
\includegraphics[width=8cm]{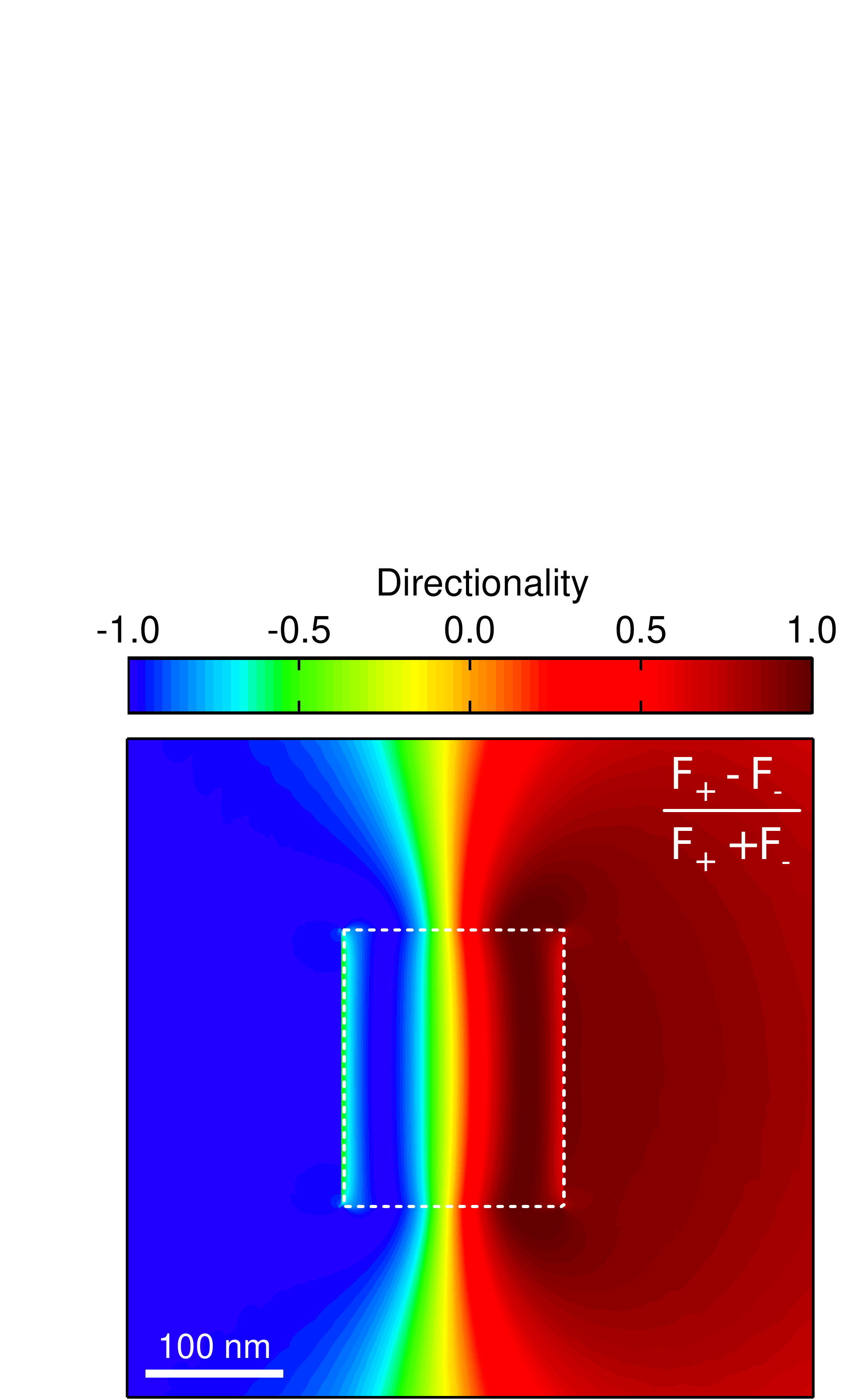}
\caption{The directionality of emission for a right-handed dipole $\mathbf{u}_+$ in a ring resonator, derived from the results presented in Fig.~\ref{fig:f2_VnF}.  For this ring the directionality ranges from +1 (all emission into the right-handed mode; outer side of the resonator) to -1 (all emission into the left-handed mode; inner side of the resonator).}
\label{fig:f3_directionality}
\end{center}
\end{figure*}

The directional nature of the emission of the circular dipole is clear from Fig.~\ref{fig:f2_VnF} since regions of high $F_+$ do not overlap with regions of high $F_-$.  We quantify this directionality in Fig.~\ref{fig:f3_directionality}, where we plot the normalised difference of the rate of emission between $\Gamma_+$ and $\Gamma_-$ for a right-handed circular dipole $\mathbf{u}_+$.
Here, we observe that perfect directional emission is possible for this small ring resonator with regions of +1 directionality located inside the resonator on its outer side, while regions of where the directionality is -1 can be found both inside and outside the ring in its inner side.  As before, this asymmetry is due to the bending of the waveguide, which pushes the fields to the outer side of the ring.

\begin{figure}
\begin{center}
\includegraphics[width=8cm]{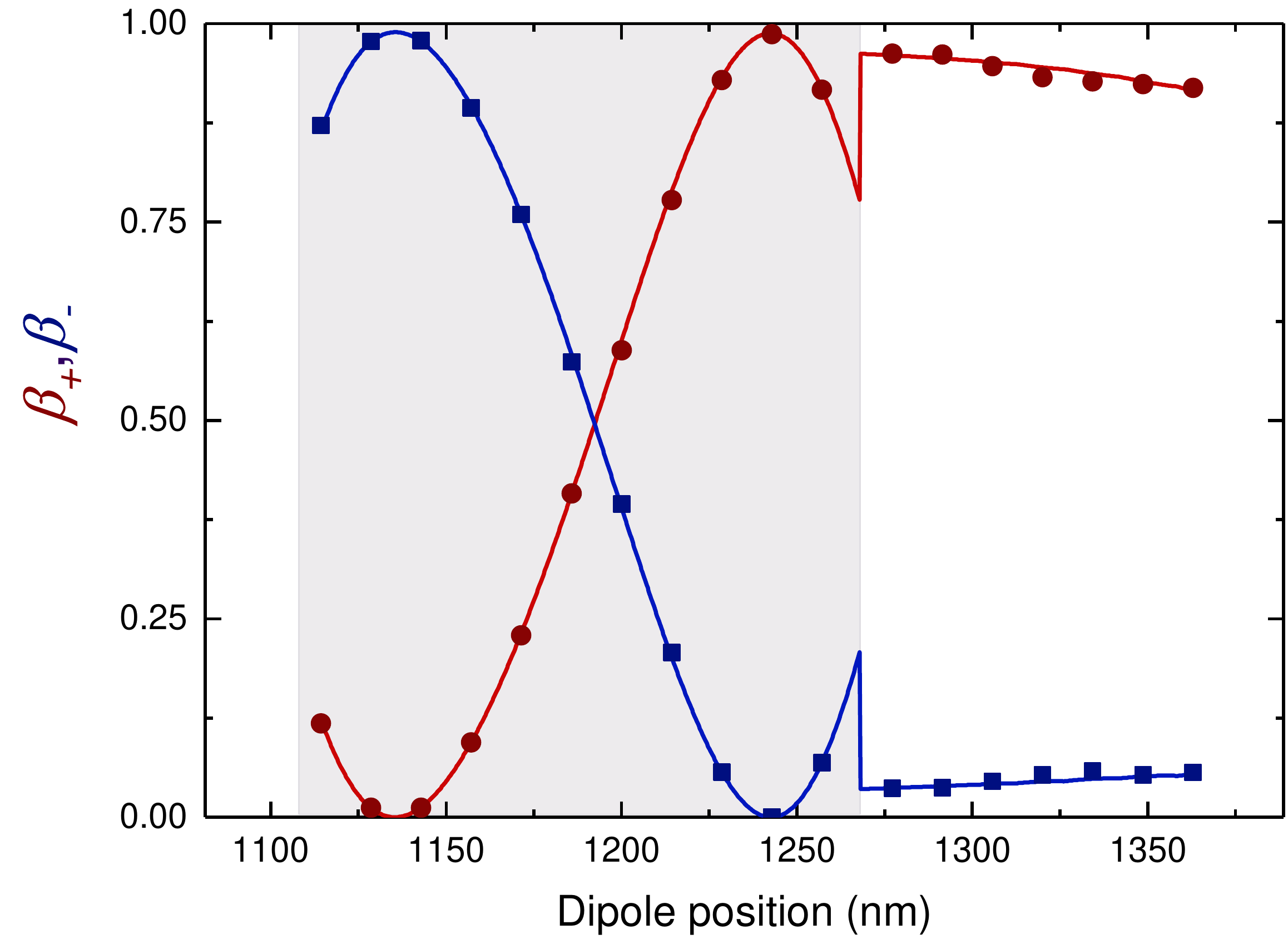}
\caption{(color online) Directional coupling efficiencies of a $\mathbf{u}_+$ dipole into the nanophotonic ring resonator, as a function of dipole position along the symmetry axis of the waveguide (c.f. Fig.~\ref{fig:f2_VnF}).  The symbols are the $\beta_\pm$ extracted from full-vectorial three-dimensional calculations, while the solid curves are found using Eqs.~\ref{eq:beta} \ref{eq:chiralF} and \ref{eq:chiralV}, and a single, two-dimensional eigenmode calculation.  The shaded region denotes the waveguide.}
\label{fig:f4_beta}
\end{center}
\end{figure}

Our theory can also be used to quantify the directional coupling efficiencies $\beta_{\pm}$, which determine how many photons are emitted into each mode of the ring. Importantly, as we discuss below, our approach allows for the calculation of $\beta_{\pm}$ without resorting to power-flux monitors, which greatly complicate the calculations in the presence of large radiation losses or absorption.  To begin with, however, when $\gamma$ is negligible, as is the case for our ring resonator, we can immediately use Eq.~\ref{eq:beta}. We plot $\beta_\pm$ along the symmetry axis of the waveguide (solid lines in Fig.~\ref{fig:f2_VnF}) in Fig.~\ref{fig:f4_beta}, observing excellent agreement between our theory (curves) using our single, eigenmode calculation with Eq.~\ref{eq:chiralF} and full-vectorial three-dimensional calculations (symbols, see below for explanation).
For this ring, both $\beta_+$ and $\beta_-$ assume peak values of $\approx 0.99$  inside the waveguide and, in the region that we consider, a maximum value of $\beta_+ \approx 0.96 $ occurs just to the outside.  Conversely, we also observe a crossing point near $r\approx1200$ nm, where $\beta_+ = \beta_-$ and hence the emission is bidirectional.

If other significant decay channels are available to the emitter, for example through radiation into free-space, then $\gamma$ must also be quantified if $\beta_{\pm}$ are to be determined.\bibnote{In the case absorption losses, the non-radiative component of each emission rate, $\Gamma_{\mathrm{nrad}} = \Gamma_\pm - \Gamma_{\pm, \mathrm{rad}}$ can be simply calculated\cite{Sau13} from the eigenmode and the absorption volume integral
\begin{equation}\label{eq:nonRad}
  \Gamma_{\mathrm{nrad}} = \frac{2}{\hbar}\int d^{3}\mathbf{r}\mbox{Im}\left(\epsilon\right) \left| \alpha_+{\mathbf{\tilde{E}_+\left(r\right)}}+ \alpha_-{\mathbf{\tilde{E}_-\left(r\right)}} \right|^2
\end{equation}}
In practise this can be done in two ways, using the three-dimensional simulations:\cite{Rot17} (i) Either, one calculates $\beta_{\mathrm{tot}}$, by fitting from the leaky mode envelope, which then allows for the calculation of $\beta_\pm = \beta_{\mathrm{tot}} F_\pm / \left(F_+ + F_-\right)$, where $F_\pm$ is found using Eq.~\ref{eq:chiralF}. (ii) Alternatively, the total emission enhancement $\eta$, which is defined to be the ratio of the power radiated by the dipole in the resonator to that of the dipole in bulk, and which in light of Eq.~\ref{eq:beta}, is related to the directional coupling efficiencies through $\beta_\pm = F_\pm / \eta$.  Approach (ii), in particular, does not require any fitting and provides a clear and unambiguous way to calculate the directional emission efficiencies numerically.

\section{Conclusions and outlook}
The ability to accurately and efficiently quantify the interactions of quantum emitters with nanophotonic structures is key if the promise of theoretical proposals and proof-of-concept experiments is to be fulfilled.  For chiral quantum interactions, we have provided a framework within which quantum electrodynamic parameters such as directional decay rates and coupling strengths, can be calculated for realistic and compatible photonic resonators.  As an example, we have calculated these parameters for a dielectric ring resonator with computationally favorable 2D models, and observed excellent agreement between our theory and full-numerical 3D simulations.  We have shown that, even with this simple resonator, careful positioning of an emitter allows for any degree of directionality, all with a high Purcell enhancement and near-unity coupling efficiency.  This demonstrates the potential of our approach to design and engineer realistic and efficient quantum chiral interfaces, representing an important step towards the realization of quantum photonic processing~\cite{Pic15, Mah16, Lod17}. Our theory also describes classical chiral light-matter interactions~\cite{Pet14, Feb15} and could therefore open new routes towards to create and design chiral photonic elements, or to enhance the spectroscopy and sensing of chiral molecules \cite{pat13}.

\begin{acknowledgement}

The authors acknowledge support from the Max Planck Society and thank Pierre T\"{u}rschmann and Peter Lodahl for stimulating discussions.

\end{acknowledgement}

%
%




\begin{mcitethebibliography}{32}
\bibitem[Not()]{Note-1}
Here, the real part of $\tilde{\omega}$ is the resonance frequency while the
  imaginary component quantifies the damping of the resonator.\relax
\mciteBstWouldAddEndPunctfalse
\mciteSetBstMidEndSepPunct{\mcitedefaultmidpunct}
{}{\mcitedefaultseppunct}\relax
\EndOfBibitem
\bibitem[Not()]{Note-2}
Note that the same is true for the ellipse of the $E_z$ and $E_\varphi$
  components, and for the magnetic components, where sign of the longitudinal
  component flips\relax
\mciteBstWouldAddEndPuncttrue
\mciteSetBstMidEndSepPunct{\mcitedefaultmidpunct}
{\mcitedefaultendpunct}{\mcitedefaultseppunct}\relax
\EndOfBibitem
\bibitem[Not()]{Note-3}
Note that a generalization to dispersive resonators is nevertheless
  possible\cite{Sau13}\relax
\mciteBstWouldAddEndPuncttrue
\mciteSetBstMidEndSepPunct{\mcitedefaultmidpunct}
{\mcitedefaultendpunct}{\mcitedefaultseppunct}\relax
\EndOfBibitem
\bibitem[Not()]{Note-4}
Assuming that the modes do not spectrally overlap, we need only consider the
  two counterpropagating modes so that the sum runs over $\{+,-\}$ and we omit
  the frequency index for brevity; the calculation can be repeated for each
  pair of degenerate QNMs.\relax
\mciteBstWouldAddEndPunctfalse
\mciteSetBstMidEndSepPunct{\mcitedefaultmidpunct}
{}{\mcitedefaultseppunct}\relax
\EndOfBibitem
\bibitem[Not()]{Note-5}
In the case absorption losses, the non-radiative component of each emission
  rate, $\Gamma_{\mathrm{nrad}} = \Gamma_\pm - \Gamma_{\pm, \mathrm{rad}}$ can
  be simply calculated\cite{Sau13} from the eigenmode and the absorption volume
  integral \begin{equation}\label{eq:nonRad} \Gamma_{\mathrm{nrad}} =
  \frac{2}{\hbar}\int d^{3}\mathbf{r}\mbox{Im}\left(\epsilon\right) \left|
  \alpha_+{\mathbf{\tilde{E}_+\left(r\right)}}+
  \alpha_-{\mathbf{\tilde{E}_-\left(r\right)}} \right|^2 \end{equation}\relax
\mciteBstWouldAddEndPuncttrue
\mciteSetBstMidEndSepPunct{\mcitedefaultmidpunct}
{\mcitedefaultendpunct}{\mcitedefaultseppunct}\relax
\EndOfBibitem
\end{mcitethebibliography}
\end{document}